\documentstyle[12pt]{article}
\newcommand{\be}{\begin{equation}}
\newcommand{\bef}{\begin{figure}}
\newcommand{\eef}{\end{figure}}

\newcommand{\ee}{\end{equation}}
\def\spose#1{\hbox to 0pt{#1\hss}}
\def\ltapprox{\mathrel{\spose{\lower
3pt\hbox{$\mathchar"218$}}
 \raise 2.0pt\hbox{$\mathchar"13C$}}}
\def\gtapprox{\mathrel{\spose{\lower
3pt\hbox{$\mathchar"218$}}
\raise 2.0pt\hbox{$\mathchar"13E$}}}
\def\inapprox{\mathrel{\spose{\lower
3pt\hbox{$\mathchar"218$}}
 \raise 2.0pt\hbox{$\mathchar"232$}}}

\begin{document}
\title{\bf Field Theory of Gravitation: Desire and Reality}

\author{Yurij V. Baryshev \\ \\
\small Astronomical  Institute of the Saint-Petersburg University, St.-Petersburg, Russia. \\
\small Scientific-Educational Union "Earth and Universe", St.-Petersburg, Russia.\\
\small E-mail:  yuba@astro.spbu.ru}

\date{}

\maketitle

%\centerline{}
%\centerline{}

\begin{abstract}
A retrospective analysis of the field theory 
of gravitation, describing
gravitational field in the same way as other fields of matter in
the flat space-time, is done. The field approach could be called
"quantum gravidynamics" to distinguish it from the "geometrodynamics"
or general relativity. The basic propositions and
main conclusions of the field approach are discussed with
reference to classical works of Birkhoff, Moshinsky, Thirring,
Kalman, Feynman, Weinberg, Deser. In the case of weak fields both
"gravidynamics" and "geometrodynamics" give the same
predictions for classical relativistic effects.  However, in the
case of strong field, and taking into account quantum nature of
the gravitational interaction, they are profoundly different.
Contents of the paper:
1) Introduction;
2) Two ways in gravity theory:
2.1.Hypotheses of Poincar\'e and Einstein,
2.2. Gravity as a geometry of space,
2.3. Gravitation as a material field in flat space-time;
3) Classical theory of tensor field:
3.1.Works of Birkhoff and Moshinsky,
3.2.Works of Thirring and Kalman,
3.3.Thirring and Deser  about identity of GR and FTG;
4) Quantum theory of tensor field;
5) Modern problems in field theory of gravitation:
5.1.Multicomponent nature of tensor field,
5.2.Choice of energy-momentum tensor of gravitational field,
5.3.Absence of black holes in FTG,
5.4.Astrophysical tests of FTG;
6) Conclusions.

\end{abstract}

\section{Introduction}

There is a common statement both in scientific publications and
popular literature dealing with General Relativity (GR) that
geometrical description of gravity is the only
logically consistent
generalization of the Newtonian classical theory of gravitation.
However, a reader, non-aligned to
general relativity may put a natural question why 
it is impossible to consider gravitation in the same way as other
physical interactions, i.e.  as a quantum field in flat
space-time background.

Indeed, such a field approach to gravity has been discussed 
in the literature and known since the works of Poincar\'e in 1905-1906
on the special theory of relativity. The Field Theory of Gravitation
(FTG) was considered in classical works of Birkhoff, Moshinsky,
Thirring, Kalman, Feynman, Weinberg, and Deser. The history
of FTG is full of misleading claims and it demonstrates
the hard way of creation and development of scientific ideas.

There are many articles and books dealing with GR but only
a few papers discuss FTG. Perhaps it is a consequence
of wide-spread opinion that FTG is equivalent to GR and hence
we need not spend time to study field gravity approach.
This opinion comes from the basic textbooks on GR, such as
Misner, Thorn, Wheeler (1973), ch. 18, and Zel'dovich, Novikov
(1971), ch. 2, where it is asserted a possibility
for derivation of GR from tensor field theory of gravitation
and hence the same physical meaning of both theories.
Indeed in papers of Thirring(1961) and Deser (1970) there were 
claims that field theory approach is identical with the
geometrical one and there are no gravitational effects which
could provide grounds to distinguish between them.

Certainly, this natural desire to
justify geometrisation via field theory approach 
(i.e. utilizing common basis with other physical interactions)
is very clear. However, as it will be shown here, 
reality turns out to be much more
complex and interesting. Actually GR and FTG are two alternative
theories with different basises  and  different
observational predictions. Of course, for the weak gravitational fields, 
which are available for experiments now,
both theories give the same values
of the classical relativistic effects,
but they profoundly different in the case of strong gravity, which will be
obtainable in near future. 
Astrophysical observations of double pulsars,
massive compact objects in double X-ray sources (the so called
"black holes candidates"), active galaxy nuclei,
gravitational waves and large scale distribution of matter in
the Universe will help to check these theories in strong gravity
limit.

In the present paper
we shall discuss main postulates, results and problems 
of field gravity theory in historical perspective 
and in comparison with general relativity.

\section{ Two ways in gravity theory }

\subsection{ Hypotheses of Poincar\'e and Einstein}

As early as 1905 Henri Poincar\'e in his work
"On the Dynamics of the Electron"
 put forward an idea to build
relativistic theory for all physical interactions, including
gravitation, in flat four-dimensional space . He pointed out that
gravitation should propagate with the velocity of light in a
retarded way --- analogous to electrodynamics, and there should
exist interaction mediators --- gravitational waves.
Several years later in his lecture on "New conceptions of matter"
Poincar\'e wrote about inclusion of Planck's discovery of
quantum nature of electromagnetic radiation in the framework of
future physics.  Pioncar\'e, thus, could be rightfully regarded as
the founder of that direction in theory of gravitation which now
is called the quantum theory of gravitational interaction
in flat space-time. This way is analogous to that
of all non-gravitational physics and it came to foundations of
such basic theories as quantum electrodynamics, quantum theory
of weak interactions, and quantum chromodynamics. Naturally, the
quantum gravidynamics or the field theory of gravitation  -
should take its place in this line.

In 1915 Albert Einstein published his basic equations of general
relativity  and thus opened another way for
gravitational theory. GR treats the gravity as a curvature of
the space itself caused by all non-gravitational matter, but not
as a kind of material field distributed in space. Wheeler called
this approach later as geometrodynamics. According to Einstein,
the gravitational interaction stands aside comparing with other
physical ones for which space-time is a passive background. The
gravitational force is caused not by material interaction
carriers, but by curvature of space, i.e. its deviation from
Euclidian geometry. In fact, this means "the materialization of
space", because the space can be deformed, expand and even
spread in the form of gravity waves.

Thus, at the beginning of 20th century there formed two
alternative ways for the theory of gravitation. The way of
Poincar\'e --- is the description of gravity as a relativistic
quantum field in flat space-time, the unique approach to all
physical fields. The way of Einstein --- is the reduction of
gravity to space-time curvature, i.e. the statement of gravity
as an exclusive among others interaction.

\subsection{Gravity as a geometry of space}

The overwhelming majority of works in gravity theory
follow the way opened by Einstein. At the beginning of its
development, the geometrical description of gravity was strongly
motivated by the {\it principle of equivalence}. 
However as it was emphasized by Fock,
strictly speaking it is not a physical but a "philosophical"
principle. Indeed, there are many variants of its formulation
but they cannot be tested experimentally. 
The real background of geometrical
approach is the so called {\it principle of geometrisation},
according to which all gravitational phenomena can be described
by the metric of the Riemann space
$g^{ik}$.
Modern achievements of GR are
presented in a collection of review articles "Three hundred
years of gravitation" (1987) edited by Hawking and Israel.

It should be noted that GR is a mathematically precise
non-linear theory without any inner limitations to its physical
applications. Thus the solutions of the Einstein's equations  are
considered to be physically valid both inside and outside the
gravitational radius 
$R_g = 2GM/c^2$,
end even up to infinite densities in singularity.
Moreover geometrical approach yields the infinite gravitational
force at the finite radius
$R_g$.
The infinite force cannot be balanced by any finite
(non-gravitational) force and this leads to body's collapse
to singularity, the so called "black hole". GR inevitably comes
to existence of such exotic objects (strictly speaking
mathematical ones) as black holes with infinite forces, time
machines with junction between past and future, expanding
universes with continuous creation of space.

The prediction of "black holes" is believed to be one of the
great achievements of GR, while the problem of energy of
gravitational field is considered as its "Ahill's heel" for more
then 80~years. This problem is a permanent point of discussion
in literature since the work of Schr\"odinger in 1918 in which
he showed that the energy-momentum complex in GR is not a tensor
(i.e.~it has no definite physical value).

The roots of this problem are connected with backgrounds of
geometrical approach. Actually, the flat space guaranties the
conservation laws of energy, momentum and angular momentum
according to Noether's theorem. This problems has been under
discussion in recent years by Logunov with  collaborators
(see his book "Lectures on relativity theory and gravitation").
Rejection of flat space inevitably leads to fundamental
difficulties with energy conservation. The famous statement from
the textbook of Landau and Lifshitz (1971), ch. 11, sec. 96,
is that general covariant form of energy conservation 
\be
\label{1}
T_{\phantom{k}i;k}^{k} = 0
\ee
actually "does not express any conservation law whatever",
and it presents just brilliant form of energy problem in GR.
Moreover, the infinite gravitational force
acting at the gravitational  radius of
a "black hole" require infinite energy of gravitational field
(of course if it exists).

Among the "standard solutions" of the energy problem in GR there are
such statements as the absence of the "old" concept of energy
of gravitational field, or that the gravitational energy
is a non-localizable quantity. If the former
takes place, it is unclear why so much efforts are devoted to
build gravity wave detectors, the devices just aimed to
localize gravitational energy. Non-local energy cannot be
treated in quantum way, and this is a reason why there are
no geometrical quantum theory of gravitation.
Also this is why there are attempts to construct
in the frame of GR the true energy-momentum tensor
of gravity "field" (see e.g. Babak\&Grishchuk,1999).
However the geometrical approach (by definition, via the
principle of equivalence) excludes both gravity force and
localaizable energy of gravity. Hence the only way to construct
the true EMT of gravity field is actually to develop
the true field approach as discussed below.

\subsection{ Gravitation as a material field in flat space-time}

Following the way which was pointed out by Poincar\'e,
the classical relativistic field gravity theory has been developed 
by Birkhoff(1944), Moshinsky(1950), Thirring(1961), and Kalman(1961).
A quantum extension of FTG was considered by Feynman(1963) 
and Weinberg(1965).
The basis of FTG is the Lagrangian formalism of relativistic field 
theory, which is applied to the symmetric tensor of second rank
$\psi^{ik}$
representing the gravitational field in Minkowski space. 

The principle of equivalence can not be a foundation of the field
approach because this principle eliminates gravity force and
state equivalence between inertial motion and accelerated motion
under gravity ("natural state of free folling"). For instance this
principle creates such puzzl as the radiation of an electric charge
resting on a laboratory table in the Earth's gravity field due to
equivalence of this frame to the constant acceleration of the table.

The basic concept of inertial frame is conserved in FTG and there is
no equivalence between inertial and accelerated motion. Instead of
principle of equivalence there is the 
{\it principle of universality} of gravitational
interaction (see below Eq.10), which has explicite relativistic form.
As a consequence of this principle
all bodies fall with the same acceleration in the Earth's 
gravity field and so the principle of equivalence in this form
is satisfied. Moreover all classical relativistic effects
in weak gravitational field have the same values in both FTG and GR.

The concepts of gravity force and gravity field energy are
tightly connected in FTG and they are relativistic quantum
physical quantities.
In FTG the energy is considered as primary and fundamental
concept, at least for the reasons of quantum approach where one
needs the energy of gravitational field for its quantization.
The flat space background without curvature, expansion and
contraction (i.e. without creation and disappearance of vacuum)
plays a very important role, radically changing the gravitational 
theory itself.

In flat space-time  Noether's theorem leads to conservation laws for
energy, momentum and angular momentum, including gravitational
field. Instead of Eq.\ref{1} we get in flat space
\be
\label{2}
T_{i,k}^{k} = 0,
\ee
with ordinary partial derivative in place of covariant one.  In
FTG Eq.\ref{2} gives both conservation laws and equations of motion; in
GR Eq.\ref{1} gives only the equations of motion.

The existence of the energy-momentum tensor (EMT) of
gravitational field in its traditional form allows for the
ordinary quantization of gravitational field in FTG. The
resulting quanta --- both real and virtual carry gravitational
interaction between physical bodies and fill physical vacuum ---
flat Minkowski space-time. The local positive energy density of
gravitational field excludes the existence of such hypothetical
objects as black hole and time machines. This statement can
easily be proved and has clear physical meaning (see
section 5.3).

It should be stressed that Lagrangian formalism doesn't allow to
get a uniquely defined form of EMT for any field. If
$T^{ik}$
is a EMT of the field, then any new tensor of the form
\be
\label{3}
\hat T^{ik} = T^{ik} + \frac{\partial}{\partial x^n}\,
\psi^{ikn},
\ee
with arbitrary third rank tensor
$\psi^{ikn}$
satisfying the antisymmetry relation
\be
\label{4}
\psi^{ikn} = - \psi^{ink},
\ee
is also another EMT for the same field.

In the book of Bogolyubov and Shirkov (1976), ch. 12,
sec. 2, authors remark that this ambiguity in EMT definition
is the main reason why they do not consider the gravity theory.
The main point is that EMT plays a role of the field source and
should be uniquely defined, which requires to use
additional physical conditions. For instance, there are
following natural conditions for EMT:  symmetry, positive field
energy, zero trace for massless fields.  All these conditions
are fulfilled for electromagnetic field. In following chapters
we discuss the problems arising in case of gravitation.

Note that there is another approach to gravity theory which try
to connect flat Minkowski space and curved Riemann space by
postulating the {\it principle of geometrisation}. According to this
principle all kinds of matter moves in effective Riemann space,
while Minkowski space actually exist. Such approach is called
"Relativistic Theory of Gravitation" (RTG) and is developed by
A.A.Logunov with collaborators (see e.g. Logunov (1987);
Logunov, Mestvirishvili (1989); Genk (1995); Vyblyi (1996)).

The main difference between FTG and RTG is that the field
approach does not use any geometrisation, but is based on the
concept of quantum symmetrical tensor field which has positive
localizable energy density (for static and free field)
corresponding to positive energy of gravitons --- mediators of
gravitational interaction. The geometrical picture appears in
FTG only as approximation in some classical problems.
Below we shall consider the history and main results of the 
field approach to gravity.

\section{ Classical theory of tensor field}

\subsection{ Works of Birkhoff and Moshinsky}

Forty years passed since Poincar\'e for the first time had put
forward an idea to build relativistic theory of gravitation
before the first real step was made in its fulfillment. It was
done by Birkhoff in his work "Flat space-time and
gravitation" (1944) where he formulated the theory "independent
of all ideas of curved space-time and of the
corresponding Einstein's theory".

It was a  phenomenological theory of symmetric tensor field
$\psi^{ik}$
in flat Minkowski space with metric
$\eta^{ik}$.
Birkhoff postulated the equations of field and motion in
given field in the form:
\be
\label{5}
\Box\psi^{ik} = \frac{8 \pi G}{c^2}\, T^{ik},
\ee
\be
\label{6}
\frac{dp_i}{ds} = - mc
(\psi_{ik,m} - \psi_{km,i})
u^k u^m,
\ee
where
$\Box$
wave operator,
$T^{ik}$ is
EMT of the field sources,
$u^i$
--- 4-velocity and
$p^i= mcu^i$
--- 4-momentum of the test particle with the rest mass
$m$
which drops out from both parts of Eq.\ref{6} (from here  the
notations of Landau and Lifshitz (1971) are used). All risings
and lowerings of indexes are ordinary made with Minkowski
tensor
$\eta^{ik}$,
for instance
\be
\label{7}
\psi_{i}^{k} = \eta^{km} \psi_{im}\;.
\ee

Birkhgoff noticed that if for gravitational potential we use the
tensor (now known as "Birkhgoff potential"):
\be
\label{8}
\psi^{ik} = \varphi_N \; {\rm diag} (1, 1, 1, 1)\;,
\ee
with
$\varphi_N$
being classical Newtonian potential, we arrive to all known
relativistic effects --- light bending, gravitational frequency
shift, advance of Mercury's perihelion. The price of it is the
adoption that EMT of matter should be of the form
\be
\label{9}
T^{ik} = \frac{1}{2} \rho c^2\; {\rm diag} (1, 1, 1, 1)\;,
\ee
i.e., he postulated the super-hard equation of state
$\varepsilon = p$
with
$^1\!/\!_2$
as additional multiplier.

Obviously, such postulate and the absence of derivation of
his equations from more general propositions
were weak points of the suggested theory. 
This is why Weyl interpreted this
fact in terms of impossibility to build consistent field theory
of gravitation and inevitability of geometrical approach.

Next important step in building the FTG was made by Moshinsky in
his work "On the interaction of Birkhoff's gravitational field
with the electromagnetic and pair fields" (1950). He for the
fist time made detailed calculations of the main relativistic
effects consistently using the interaction Lagrangian in the
form:
\be
\label{10}
\Lambda_{<int>}= - \frac{1}{c^2} \psi_{ik} T^{ik}.
\ee

In fact, Moshinsky was the first who formulated the
{\it principle of universality} of the gravitational
interaction (Eq.\ref{10}), according to which all kinds of matter
interact with gravitational field through its EMT. This
principle is a key stone in field approach to gravitation and
plays the same fundamental role as the principle of
geometrisation does in geometrical description of gravity.

Starting from Birkhgoff's potential (Eq.\ref{8}) and the principle of
universality (Eq.\ref{10}) Moshinsky accurately calculated the effects
of light bending, hydrogen atom energy levels shifts and
gravitational corrections to electron's magnetic moment.
However, at that time there was no satisfactory derivation of
Birkhgoff's potential itself and Moshinsky's paper passed
practically unnoticed.

\subsection{ Works of Thirring and Kalman}

First full and consistent formulation of the FTG as a theory of
symmetric tensor field in Minkowski space was published in
Thirring's paper "An alternative approach to the theory of
gravitation" (1961).  In the weak field approximation Thirring
obtained the non-linear equations:
\be
\label{11}
\Box \psi^{ik}\;=\; \frac{8 \pi G}{c^2}
\left(
T^{ik} -\;\frac{1}{2} \eta^{ik} T
\right)\!;
\ee
\be
\label{12}
T^{ik}\;=\,T_{<m>}^{ik}+\;T_{<int>}^{ik}+\;T_{<g>}^{ik},
\ee
where
$T^{ik}$
is the total EMT of the system, including gravitational field
itself,
$T_{<m>}^{ik}$
is the EMT of matter,
$T_{<int>}^{ik}$
is the EMT of the interaction,
$T_{<g>}^{ik}$
is the EMT of the gravitational field,
$T$
is the trace of the total EMT.

The equation (\ref{11}) differs from that of Birkhgoff's Eq.\ref{5} in two
ways: first, it contains the term
$^1\!/\!_2\eta^{ik}T$,
second, there are corrections connected with the energy-momentum
of the gravitational field. Now, the Birkhgoff's potential (Eq.\ref{8})
can be accurately derived as a linear approximation of the
Thirring's equations with the full EMT being equal to that of
matter's one:
\be
\label{13}
T^{ik}=\;T_{<m>}^{ik}=\;\rho c^2 \;{\rm diag} (1, 0, 0, 0)\;,
\ee
which corresponds to ordinary "dust-like" equation of state
instead of super-hard one in original Birkhoff's formulation.
This statement removed the main
Weyl's objection against FTG.

Thirring, also for the first time, put a question about unique
definition of the EMT of gravitational field in connection with
calculations of nonlinear corrections in Eq.\ref{11}. He
showed that canonical form of gravitational field EMT possesses
symmetry and yields positive value for energy density of static
field, described with Birkhgoff's potential (Eq.\ref{8}):
\be
\label{14}
T_{<g>}^{00}\;=\; \varepsilon_{<g>}=\; \frac{1}{8 \pi G}\,
(\nabla \varphi_N)^2 \ \frac{{\rm erg}}{{\rm cm}^3},
\ee
Moreover, EMT of interaction gives a negative contribution to
the energy density (note, that
$\varphi_N < 0$)
\be
\label{15}
T_{<int>}^{00}= \rho \varphi_N,
\ee
and the full energy of gravitationally bound system turns out to
be less then the total rest energy of its constituting parts
(see the discussion about this in Sokolov S.(1995) and Baryshev
(1995b)). As we noted above, from the fact that the energy of
gravitational field is positive it follows that there are no
"black holes", but Thirring didn't notice this fact (see 5.3).

Unfortunately, an error occurred in Thirring's work --- in deriving
the equation of motion of test particles the variation of proper
interval
$ds$
were not taken into account (see the equation (11) in Thirring's
paper). This error was corrected in the work of Kalman
"Lagrangian formalism in relativistic dynamics" (1961) where he
got the equations of motion in tensor field which could be
written in the form (see also alternative derivation of this
equations in Baryshev,1986)
\be
\label{16}
A_{k}^{i} \frac{du^k}{ds}\;=\;-B_{kl}^{i} u^k u^l,
\ee
where
\be
\label{17}
A_{k}^{i} \;=\;
\left(
   1-\frac{1}{c^2} \psi_{ln} u^l u^n
\right)
{\eta}_{k}^{i}
\;-\;
\frac{2}{c^2}\psi_{kn} u^n u^i
\;+\;
\frac{2}{c^2} \psi_{k}^{i}\;,
\ee
\be
\label{18}
B_{kl}^{i}\;=\; \frac{2}{c^2} \psi_{k,l}^{i}
\;-\;
\frac{1}{c^2} \psi_{kl}^{,i}
\;-\;
\frac{1}{c^2} \psi_{kl,n} u^n u^i.
\ee

Equations (16) have the tensor form hence particles trajectories
do not depend on choice of coordinates.

In static spherically symmetric field in post-Newtonian
approximation it is enough to write the non-linear correction
only for
$\psi^{00}$
component of Birkhgoff's potential:
\be
\label{19}
\psi^{00} = \varphi_N + \frac{1}{2}
\frac{(\varphi)^2}{c^2},
\ee
where
$\varphi_N = - GM/r$
is the Newtonian potential. Substituting the Birkhgoff's
potential (Eq.\ref{8}) with correction (Eq.\ref{19}) 
to Kalman equations (16) we
arrive to the three dimensional form of equations of motion:
\be
\label{20}
\frac{d{\bf v}}{dt}
\;=\;
-\left(
   1+ \frac{v^2}{c^2} + 4 \frac{\varphi_N}{c^2}
\right)
\nabla \varphi_N
\;+\;
4 \frac{{\bf v}}{c}
\left(
   \frac{{\bf v}}{c} \cdot \nabla \varphi_N
\right)\!.
\ee

Equation (20) gives for the advance of perihelion of a test
particle:
\be
\label{21}
\delta \varphi = \frac{6 \pi GM}{c^2 a (1-e^2)},
\ee
which coincides with the known formula in GR.
Together with Moshinsky's results on electromagnetic
and spinor fields interacting with gravity field, this shows
that classical relativistic effects have the same values
in both GR and FTG.

An interesting consequence of the equation (20) is a possibility
of generalisation of the old Galileo's experiment
with free folling bodies. A new version of such Galileo-2000 experiment
is based on the fact that gravitational
acceleration and gravitational force 
acting on a test particle depend on velocity of the particle.
For example, as it was shown by Baryshev \& Sokolov (1983), 
rotating bodies with different values and orientations of angular momentum
will folling down from a "drop tower" with different accelerations.
And this is direct consequence of the principle of universality of
gravitational interaction and simply means that gravity force
depends on the velocity of test particle. Another version of this
experiment is a weighing rotating bodies with a scale, where
there is no a free folling but there is actual balance of gravity
forces acting on two differently rotating bodies.

Kalman made a comparison of linear theories for scalar, vector
and tensor fields and, for the first time, derived exact
expression of EMT for interaction between test particles and
gravitational field in linear approximation. He also got a
general form of energy-momentum conservation law for the system
consisting of particles plus field in the form:
\be
\label{22}
\left[
T_{<m>}^{ik} +\; T_{<int>}^{ik} +\; T_{<f>}^{ik}
\right]_{,k} = 0.
\ee
It is interesting that for pure vector field (electrodynamics)
the conservation law can be written as
\be
\label{23}
\left[
T_{<m>}^{ik} +\; T_{<f>}^{ik}
\right]_{,k} = 0,
\ee
i.e. without EMT of interaction (Kalman,1961).

\subsection{Thirring and Deser about identity of GR and FTG}

Characteristic feature of almost all works on FTG is that
authors try to declare the identity of field and geometrical
approaches. Thirring's article (1961) starts with: "The problem
of gravity has been solved by Einstein in his general theory
of relativity... Any further contribution in this field
is necessarily pedagogical or interpretational."  Namely the
reference to Thirring gave the foundation for Zel'dovich and
Novikov (1971) to state that field approach to gravitation is
identical with GR.

However, reader can easily find out that Thirring's work deals
only with the case of weak field where both approaches give the
same results.The situation in strong field, where they differ
profoundly, is not considered at all.

Thirring suggest to consider the sum of two tensors
$\eta^{ik}$
and
$\psi^{ik},$
which can be interpreted as an effective Riemann space metric tensor
\be
\label{24}
\hat g^{ik} = \eta^{ik} + \psi^{ik}.
\ee

It ought to be noted that this presentation is a secondary and
approximate in the framework of FTG. The prime features of FTG
are quantum gravitational field and its quanta --- gravitons, which
cannot be eliminated with any choice of reference system. From
this it follows that principle of equivalence in formulation
analogous to the GR doesn't take place in FTG, though, as was
shown by Thirring, the equality of inertial and gravitational
masses holds in FTG exactly and is a direct consequence of the
principle of universality (Eq.\ref{10}).

Moreover, the definition (24) has one more shortcoming --- it
takes us beyond the basic suggestions of tensor analysis in flat
space. Actually, if
$\hat g^{ik}$
is a tensor in flat space, then its covariant components should
have the form:
\be
\label{25}
\hat g_{ik} = \eta_{ik} + \psi_{ik},
\ee
and its mixed components and trace are as follows:
\be
\label{26}
\hat g_{k}^{i} = \delta^{i}_{k} + \psi_{k}^{i},
\ee

\be
\label{27}
\hat g_{ik} \hat g^{ik} = 4 + 2 \psi + O (\psi_{ik}^{2}).
\ee

The true metric tensor in Riemann space must satisfy the
following exact relations:
\be
\label{28}
g_{k}^{i} = \delta_{k}^{i},
\ee
\be
\label{29}
g_{ik} g^{ik} = 4,
\ee
which differ from (26) and (27). To match this two distinct
tensors are usually considered:
\be
\label{30}
\hat g^{ik} = \eta^{ik} + \psi^{ik},
\ee
\be
\tilde g_{ik} = \eta_{ik} - \psi_{ik},
\ee
which are considered as the components of one metric tensor in
Riemann space.

Let us notice that in consistent FTG all quantities with upper
and lower indexes belong to the same tensor, i.e. the
geometrical object of the flat space. Thus there are no
confusion in interpretation of co- and contravariant components
and there is always possible to introduce global Cartesian
coordinates. The existence of flat background metric radically
distincts the FTG from GR which most clearly seen in strong
gravitational fields. Thus, there no prove of identity between
FTG and GR in Thirring's work, all what was done was the
demonstration of possibility to introduce an effective Riemann
space for weak gravitational field.

Deser (1970) summed up all the preceding attempts to derive
full system of Einstein equations from linear equations for
tensor field. The essence of his approach is an iterrative
procedure with  consequent account for the EMT of gravitational
field, obtained from Lagrangian of the previous step. It starts
from free tensor field Lagrangian and postulated form of metric
tensor:
\be
\label{32}
g^{ik} = \eta^{ik} + h^{ik},
\ee
which, as was mentioned above, is a deviation from the field
theory.

The main disadvantage of iterations is that Lagrangian
formalism doesn't allow to get a unique gravitational
EMT, so there is an infinite set of different non-linear
generalizations of linear tensor field equations.

In fact, Deser showed no more then it is possible to find such
an expression of EMT which at the third iteration gives
Einstein equations and in no way this leads to the conclusion
about identity of field and geometrical approaches, as was
claimed in the book of Misner, Thorne, Weeler (1977). Moreover,
the essence of field approach suggests such a choice of
gravitational EMT that satisfies zero trace (massless graviton)
and positive energy density of gravitational field and
namely these properties should be tested first. Deser's EMT does
not satisfy these conditions. It is easy to demonstrate that
positive energy requirement leads to radical difference of field
approach from that of geometrical one (see 5.3).

\section{Quantum theory of tensor field}

Quantum field theory is a theoretical background for description
of all known physical interactions (strong, week,
electromagnetic). Thus, the deeper understanding of
electrodynamics became possible only after the quantum
electrodynamics had been build. FTG deals with gravitation in
the same way as other matter fields are descried, so there is
quite natural to suggest the existence of quantum gravidymamics,
presenting gravitation on more fundamental level then classical
relativistic theory.

The idea of physical vacuum --- the flat Minkowski space filled
with quantum fluctuations of all fields, including gravitational
one, plays an important role in gravidynamics. In quantum theory
Newton's gravitation is a result of virtual gravitons exchange
between bodies, just as Coulomb law is a result of virtual
photons exchange.

First works on quantization of tensor field appeared in 30-es
in articles of Rosenfeld (1930) and Bronshtein (1936). Different
kinds of quantum formalism were used to calculate quantum
gravity effects in works of Ivanenko\& Sokolov (1947); Gupta
(1952); Feynman (1963); Zaharov (1965); Weinberg (1965).

All these works actually dealt with the tensor symmetrical field in
Minkowski space. Feynman(1971) in his "Lectures on Gravitation"
emphasized that "The geometrical interpretation is not really
necessary or essential to physics". 
The whole spirit and formalism used in these
works refer them totally to the field theory of gravitation,
though authors always speak about general relativity. In fact,
the only feature they use from geometrical approach is a formal
expression of Lagrangian and they give its expansion for the
case of weak field.

The main problems of quantum gravidynamics are the physically
grounded choice of Lagrangian and choice of gravitational EMT
which cannot be fixed with Lagrangian. Still, there is an
unresolved problem with quantum-field divergences. Another
difficulty is connected with the absence of experiments
on quantum gravity, due to smallness of expected
effects, which makes it much harder to proceed the theory. Let us
note, in this connection, the recently suggested experiment for
measurment of the frequency dependence of gravitational  bending 
of light by
planets (Baryshev, Raikov (1995); Baryshev, Gubanov, Raikov (1996)),
which could detect a quantum effects even in weak gravity field.
An important possibility of non-zero rest mass of the graviton
has been considered by Visser(1998).
A possibility of existance a scalar component of gravitational
field was discussed recently by Damour(1999).

\section{Modern problems in field theory of gravitation}

At the end of this review we list the most important, to our
point,  problems of field approach to gravitation. Their
solution in near future could make a new step in understanding
the physics of gravitational interaction.

\subsection{Multicomponent tensor field}

Quantum field theory requires to take into account both real and
virtual gravitons. In general case, there are 10 independent
components in symmetrical tensor field of the second rank
$\psi^{ik}$
and it is a superposition of particles with spins
$2, 1, 0,$ and $0'$
which are involved in virtual processes
\be
\label{33}
\{ \psi^{ik} \} = \{2 \} \oplus \{1 \}
\oplus \{0 \} \oplus \{0' \}.
\ee

The gauge invariance of the gravitational field equations gives
4 conditions, eliminating the particles with
spins 1 and $0'$ and leaving only 6 components, with spins 2 and 0.
 At the same time, the field source also consists of two
parts
\be
\label{34}
\{ \psi^{ik} \} = \{2 \} \oplus \{0 \} \leftrightarrow
\{T^{ik} \} = \{2 \} \oplus \{0 \}.
\ee
In particular, the Birkhgoff potential (8) can be presented as a
superposition of pure tensor (spin 2) and scalar (spin 0) parts
\be
\label{35}
\psi^{ik}
= \psi_{<2>}\!^{ik} + \psi_{<0>}^{ik}
= \varphi^{ik}+ \frac{1}{4}\, \eta^{ik} \psi
\;=\; \varphi_N {\rm diag}
\left(
\frac{3}{2}, \frac{1}{2}, \frac{1}{2}, \frac{1}{2}
\right)
+\; (-2 \varphi_N) \;\frac{1}{4} \eta^{ik}.
\ee

Substituting (35) in equations of motion of the test particle
(16) we arrive to conclusion that the total gravitational force,
acting on the particle with rest mass
$m$
is the sum of attraction force (spin 2) and repulsion force (spin~0),
which gives us the observed Newton force of gravity:
\be
\label{36}
F \;=\; F_{<2>} + F_{<0>} \;=\; - \frac{3}{2}m \nabla
\varphi_N + \frac{1}{2}m \nabla \varphi_N
\;=\; -m \nabla \varphi_N \;=\; F_N.
\ee

Thus, the FTG is, strictly speaking, the scalar-tensor theory,
already containing the scalar part in an initial tensor's trace
(in contrast to the Brance-Dikke theory which requires additional
scalar). The division of tensor potential into
components is a new physical problem and it requires additional
investigation.

\subsection{Choice of energy-momentum tensor of gravitational field}

An important problem of the choice of such gravitational EMT
which  satisfies the physical requirements of positive energy and
zero trace arise, since Lagrange formalism does not allow to fix
it in unique form. The multicomponent character of
gravitational field contributes additional difficulties. The
total Lagrangian is composed of attraction (spin~2) and
repulsion (spin~0) terms contributing to the canonical form of
the EMT with different signs and, thus, leaving the question
about the sign of the energy of free particles opened. The
division of gravitational EMT into two parts corresponding to
spins~2 and~0 and having positive energy was for the first time
made by Sokolov, Baryshev (1980). Further discussion and
different forms of EMT can be found in works Baryshev (1982);
Baryshev\& Sokolov (1983); Baryshev (1988); Baryshev (1990);
Sokolov (1992a, b, c, d); Baryshev (1995a).

There are two relativistic effects known until now, which could
be used to set restrictions on the form of EMT --- the
pericenter advance in two-body problem and gravitational
radiation. Both effects can be accurately measured in double
pulsar ${\rm PSR} 1913+16$.
Shift of pericenter indicates that energy
density of static gravitational field coincides with the value
given by Eq.14 with parts of percent accuracy. The observed
energy loss is consistent with radiation of quadruple waves
and fixes the value of
$T^{00}$
component with spin 2.

The most accurate observations of $PSR1913+16$ (Taylor et al.,1992)
revealed the existence of about
1\% excess in energy losses which could be interpreted as a scalar
gravitational radiation with spin 0. Different forms of EMT give
different values of the expected excess: 3\% Baryshev (1982); 2.2\%
Sokolov (1992a); 0.735\% Baryshev (1995a). For the final decision
there are required improvement of measurements accuracy and
account for possible non-gravity effects.

\subsection{Absence of black holes in FTG}

It is a fact of history, that Einstein and Eddington were
opponents to the existence of black holes and suspend the
development of the corresponding theory for almost 30 years
(see Bernstein,1996).
Einstein's argument against black holes was very simple.
Since Laplace it was known that for the body with radius less then
$R_g$
escape velocity exceeds that of light. 
Einstein (1939) actually inverted the Laplace's argument 
when he noticed that
in this case the velocity of a test particle falling on such 
a body also
exceeds the speed of light and this contradicts the special
relativity theory. Einstein considered the gravitational radius
as the limiting size of any physical body. 

Later it was shown that in GR this argument doesn't hold, 
because inside
$R_g$
the space is not static. However, in FTG the space is everywhere
flat and static, so the Einstein's argument is valid.

Eddington in his famous discussion with Chandrasekhar about the
fate of white dwarfs with masses over the critical one, stated
that there should be the law of nature preventing the
contraction of massive stars under their gravitational radius.

It is easy to show that there is such a law in FTG and it is the
law of conservation of energy! Indeed, if the energy density of
static spherically symmetrical field of the body with mass
$M$
and radius
$R_0$
is given by
$T^{00}$
component of the gravitational EMT (14), then the full energy
within the field around the body is
\be
\label{37}
E_{<g>} = \int_{R_0}^{\infty} \varepsilon_{<g>}
\,dV = \frac{GM^2}{2 R_0} \ {\rm erg}.
\ee

It follows from here that there is a natural limit of
contraction of a body and it is the condition that the
energy of the field should be less than that of the rest-mass
energy of the body:
\be
\label{38}
{\rm from}\quad E_{<g>} < Mc^2 \quad{\rm follows}\quad R_0 > \frac{GM}{2c^2}.
\ee

Thus, the black holes are prohibited by the energy
conservation.
This statement is precisely analogous to that of the classical
radius of electron
$R_e > e^2 /m_ec^2,$
following from requirement that field energy
$E_{<el>}$
should be less than
that of electron rest mass energy
$m_ec^2$.

Let us consider another important consequence of the positiveness
of gravitational energy density (14). Non-linear character of
gravitational equations are naturally connected with energy of
field itself and its sign give us two possible generalizations
of the Laplace equation
\be
\label{39}
\Delta \varphi = 0
\ee
for gravitational potential around the body with mass
$M$.
In FTG gravitational field energy is positive, 
localizable and dietributed around the gravitating body,
so instead of Eq.39 we get
\be
\label{40}
\Delta \varphi =+ \frac{1}{c^2} \; (\nabla \varphi)^2.
\ee
The solution of this equation with ordinary boundary conditions
is
\be
\label{41}
\varphi = - c^2\; {\rm ln} \left(1+ \frac{G M}{c^2 R} \right),
\ee
or for the force acting on unite test mass
\be
\label{42}
\frac{d \varphi}{dR}= \frac{GM}{R^2 \left( 1 +
{GM}/{c^2 R} \right)}.
\ee
From (42) it follows that under
$R \rightarrow R_m \approx GM/c^2$
gradient of potential is confined with
\be
\label{43}
g_{<max>} \leq \frac{c^4}{GM}=\frac{c^2}{R_m}
\ee
Hence according to FTG the force of gravity not only remains finite, 
but it decreases up to zero with infinitely increasing mass.

If we had negative static field energy density, as in the case
of
$T^{00}$
component of the Landau-Lifshitz pseudo-tensor
($\varepsilon_{<g>} = - 7 (\nabla \varphi)/8 \pi G$),
then we would have instead of (39)
\be
\label{44}
\Delta \varphi =- \frac{1}{c^2} \; (\nabla \varphi)^2.
\ee
with solution
\be
\label{45}
\varphi = - c^2 \;{\rm ln} \left(1- \frac{G M}{c^2 R} \right),
\ee
and force
\be
\label{46}
\frac{d \varphi}{dR}= \frac{GM}
{R^2 \left( 1 - {GM}/{c^2 R} \right)}.
\ee
In this case the force becomes infinite independently of mass
as the radius approaches to
$R \approx GM/c^2$,
just as we have in GR.

In FTG gravitational field energy is always positive and there
are no problems with infinities. These simple physical
considerations demonstrate the importance of EMT in
gravitational physics and radical change of theory
with change of the sign of field energy.

\subsection{Astrophysical tests of FTG}

Laboratory experiments can deal only with very weak gravity fields, but
there are astrophysical objects where fields are extremely
strong and where differences between geometrical and field theories
should manifest themselves clearly.

Observations of double pulsars give information about relativistic
dynamics of two body problem, including the loss of energy via 
gravitational radiation.
A detection of gravitational waves
from supernovae explosions
by means of third generation
antennas will give us a chance to establish their character
(transverse for spin 2 and longitudinal for spin 0 waves) ,
 to get restrictions on EMT of gravity field
and also to study physics of gravitational collapse
(see Baryshev,1997).

Observations of double X-ray  sources with compact massive
components will give clue to problem wether there are real
singularities or the saturation of gravitational interaction takes place. 
Frequently
one finds in literature that black holes have already been
detected, because there are systems with components more massive
than the Oppenheimer-Volkoff limit, i.e. over the three solar
masses.  This statement is not correct, since this limit exists only
in GR, but in FTG there could exist relativistic stars with
larger masses. The discussion on this subject can be found in
Baryshev, Sokolov (1984); Baryshev (1990; 1991);
Sokolov(1992a,b,c,d); Sokolov,Zharikov (1993).

Cosmology is another field of application of gravitation
theory. Present data about large scale galaxies distribution
contradict to the main point of Friedmann cosmology --- its
homogeneity. It turned out that galaxies form a fractal
structure with dimension close to 2
at least up to the distance scales bout 200 Mpc. 
This leads to a new possibilities
in cosmology (see an analysis of FTG cosmological
applications in the review of Baryshev et al., 1994).
 One of the main difference
between FTG and GR is that the field approach allows the
existence of the infinite stationary matter distribution (Baryshev,
Kovalevskij, 1990). In a stationary fractal distribution the observed
redshift has gravitational  and Doppler nature and is not connected with
space expansion as in Friedmann model.

\section{ Conclusions}

The retrospective analysis of field gravitation theory is
given and it is shown that field and geometrical approaches lead
to quite different conclusions in strong fields. Which way "is a
way to temple" will be clear from future astrophysical
observations.

The central problem of geometrical description of gravity is the
problem of "non-localizability" of the energy of
the gravitational field or uncertainty of its value. 
In contrast the
field gravity theory gives local, positive and observable value for the
energy of the gravitational field, which could be used for
quantum approach to gravitational interaction.

The field theory of gravitation is based on the principle of
universality of gravitational interaction and has some forms
of the principle of equivalence as its particular cases.
In FTG there are Minkowski background space and usual concepts
of gravity force, gravity field EMT and quanta of gravity
field - gravitons. Within FTG there is no infinite force
at gravitational radius and compact massive stars could
have masses much more than OV-limit. FTG is actually a scalar-tensor
theory and predicts existance of  tensor (spin 2) and
scalar (spin 0) gravitational waves.
Astrophysical tests of FTG will be avalable in near future.
It is quite natural that fundamental description of gravity
will be found on quantum level and geometrical description of
gravity may be considered as the classical limit of quantum 
relativistic gravity theory.

\vspace{2cm}

{\bf REFERENCES}\\
\newline

{\it Babak S., Grishchuk L. (1999)},
gr-qc/9907027.\\

{\it Baryshev Yu.V. (1982)},
Astrophysics, {\bf v.18}, p.93.\\

{\it Baryshev Yu.V. (1986)},
Vestnik LGU, Ser.1, vyp.4, p.113.  (in Russian).\\

{\it Baryshev Yu.V. (1988)},
Vestnik LGU, Ser.1, vyp.2, p.80.  (in Russian).\\

{\it Baryshev Yu.V. (1990)},
"Introduction to the tensor field theory of gravitation",
Saint-Petersburg University (unpublished Lectures).\\

{\it Baryshev Yu.V. (1991)},
in "Problems on high energy physics and field theory",
M., Nauka, p.61.\\

{\it Baryshev Yu.V. (1995a)},
in "Gravitational wave experiments", p.251, gr-qc/9911081 \\

{\it Baryshev Yu.V. (1995b)},
Gravitation, {\bf v.1}, iss.1, p.13.\\

{\it Baryshev Yu.V. (1997)},
Astrophysics, {\bf v.40}, p.224.\\

{\it Baryshev Yu.V., Gubanov A.G., Raikov A.A. (1996)},
Gravitation, 1996, {\bf v.2}, iss.1, p.72.\\

{\it Baryshev Yu.V., Kovalevskij M.A. (1990)},
Vestnik LGU, Ser.1, {\bf v.1}, p.86 (in Russian).\\

{\it Baryshev Yu.V., Raikov A.A. (1995)},
in "Problems on high energy physics and field theory",
Protvino, p.166 (in Russian).\\

{\it Baryshev Yu.V., Sokolov V.V., (1983)},
Proc. Astron.  Obeserv. LGU, {\bf v.38}, p.36 (in Russian).\\

{\it Baryshev Yu.V., Sokolov V.V., (1984)},
Astrofizika, {\bf v.21}, p.361.\\

{\it Baryshev Yu.V., Sylos Labini F.,
Montuori M., Pietroneiro L. (1994)},
Vistas in Astronomy, {\bf v.38}, p.419, astro-ph/9503074.\\

{\it Bernstein J. (1996)},
Sci.American, June 1996, p.66,\\

{\it Birkhoff G.D. (1944)},
Proc. Nat. Acad. Sci., {\bf v.30}, p.324.\\

{\it Bogolyubov N.N., Shirkov D.V., (1976)},
"Introduction to the theory of quantum fields",
M., Nauka (in Russian).\\

{\it Bronshtein M.P., (1936)},
Jour. Exper. Theor. Phys., {\bf v.6}, p.195 (in Russian).\\

{\it Damour T. (1999)}
gr-qc/9904057.\\

{\it Deser S. (1970)},
Gen. Rel. Grav., {\bf v.1}, p.9.\\

{\it Einstein A. (1939)},
Ann. Math. (Princeton), {\bf v.40}, p.922.\\

{\it Feynman R. (1963)},
Acta Phys. Pol., {\bf v.24}, p.697.\\

{\it Feynman R. (1971)},
"Lectures on gravitation",
(California Institute of Technology)\\

{\it Genk A. (1996)},
Gravitation, {\bf v.1}, iss.1, p.50.\\

{\it Gupta S. (1952)},
Proc. Phys. Soc., A65; p.161, p.608.\\

{\it Ivanenko D.D., Sokolov A.A. (1947)},
Vestnik MGU, 1947, n.8, p.103 (in Russian).\\

{\it Kalman G. (1961)},
1961, Phys. Rev., {\bf v.123}, p.384.\\

{\it Landau L.D., Lifshitz E.M. (1971)},
"The classical theory of field", Pergamon Press, N.Y.\\

{\it Logunov A.A. (1987)},
"Lectures on theory of relativity and gravitation",
Nauka, Moscow (in Russian).\\

{\it Logunov A.A., Mestvirishvili M.A. (1989)},
"Relativistic Theory of Gravitation",
Nauka, Moscow (in Russian).\\

{\it Misner C., Thorne K., Wheeler J. (1977)},
"Gravitation", W.H.Freeman \& Co., San Francisco.\\

{\it Moshinsky M. (1950)},
Phys. Rev., {\bf v.80}, p.514.\\

{\it Rosenfeld L. (1930)},
Zs. f. Phys., {\bf v.65}, p.589.\\

{\it Sokolov V.V. (1992a)},
Astroph. Sp. Sci., {\bf v.191}, p.231.\\

{\it Sokolov V.V. (1992b)},
Astroph. Sp. Sci., {\bf v.197}, p.87.\\

{\it Sokolov V.V. (1992c)},
Astroph. Sp. Sci., {\bf v.197}, p.179.\\

{\it Sokolov V.V. (1992d)},
Astroph. Sp. Sci., {\bf v.198}, p.53.\\

{\it Sokolov V.V., Baryshev Yu.V. (1980)},
Grav. and Theor.  Relat., Kazan St. Univ., {\bf v.17}, p.34.
(in Russian).\\

{\it Sokolov V.V., Zharikov S.V. (1993)},
Astroph. Sp. Sci., {\bf v.201}, p.303.\\

{\it Sokolov S.N. (1995)},
Gravitation, {\bf v.1}, iss.1, p.3.\\

{\it Taylor et al. (1992)}
Nature, {\bf v.355}, p.132.\\

{\it Thirring W. (1961)},
Ann. Phys. {\bf v.16}, p.96.\\

{\it Visser M. (1998)},
gr-qc/9705051.\\

{\it Vybli Yu.P. (1996)},
Gravitation, {\bf v.2}, iss.1, p.25.\\

{\it Weinberg S. (1965)},
Phys. Rev., {\bf v.138}, N4B, p.988.\\

{\it Zaharov V.I. (1965)},
Jour. Exper. Theor. Phys., {\bf v.48}, p.303 .\\

{\it Zel'dovich J.B., Novikov I.D. (1971)},
"Relativistic Astrophysics, v.1", Univ. Chicago Press, Chicago.\\

\end{document}